\documentclass[prd,twocolumn,superscriptaddress,floatfix,showpacs,nofootinbib]{revtex4-2}
\usepackage{amsfonts}
\usepackage{amsmath}
\usepackage{amssymb}
\usepackage{xcolor}
\usepackage{graphicx}
\usepackage[normalem]{ulem}
\def\be{\begin{equation}}
\def\ee{\end{equation}}
\def\bea{\begin{eqnarray}}
\def\eea{\end{eqnarray}}

\begin{document}

\title{Comment on ''Chiral electromagnetic surface waves on Chern-Simons interfaces'' Phys.Rev.D114(2026)L051903 and on arXiv:2606.12513 by N. Kaloper}

\author{
  \textbf{Igor Bandos$^{1,2,*}$} and \textbf{Bohdan Zvorskyi$^{3,\dagger}$}\\[0.2cm]
  \small $^1$Department of Physics and EHU Quantum Center, University of the Basque Country UPV/EHU,\\
  \small P.O. Box 644, 48080 Bilbao, Spain\\
  \small $^2$IKERBASQUE, Basque Foundation for Science, 48011, Bilbao, Spain\\
  \small $^3$Department of Physics and Technology, Kharkiv National University, Kharkiv, Ukraine\\[0.1cm]
  \small $^*$igor.bandos@ehu.eus, $^\dagger$bzvorskyy@gmail.com
}

\date{\small October 2, 2026}

\begin{abstract}
We demonstrate that Maxwell electrodynamics interacting with a planar Chern-Simons interface does not have plane wave solutions localized on the interface. We re-examine the derivation of such waves presented by N. Kaloper in [1]
and  show that
at an intermediate step
certain terms were missed in an equation therein. Taking these terms into account changes the final conclusion of [1]. We also show that the method used in [2] to reproduce the same result is not consistent.

\end{abstract}

\maketitle



\section{Introduction}

In recent papers \cite{Kaloper:2026auc,Kaloper:2026llp}, Kaloper studied the interaction of an electromagnetic field with a planar domain wall carrying a Chern-Simons term multiplied by a Heaviside step function. This dynamical system was considered as  effective field theory of a sector of a more complicated and more interesting dynamical system describing a membrane interacting with a Maxwell field and a 3-form gauge field, which was  considered previously in \cite{Kaloper:2026slg}.

The central claim of \cite{Kaloper:2026auc} was that there exist electromagnetic waves propagating parallel to the planar domain wall and localized on it, in the sense that the wave amplitudes decay exponentially with the normal distance from the interface.

Unfortunately, in our attempt to reproduce this result, we have found that such waves do not exist in this system.
In short, the reason is that Eq.~(16) of \cite{Kaloper:2026auc} (below referred to as (16)$^{ \cite{Kaloper:2026auc}}$) does not follow from Eq.~(15)$^{ \cite{Kaloper:2026auc}}$ due to discontinuity (14)$^{ \cite{Kaloper:2026auc}}$, while the  corrected  version of  Eq.~(16)$^{\cite{Kaloper:2026auc}}$, which we present in Sec. II, possesses no solutions localized on the interface.

Moreover, in  Sec. III we argue that an alternative attempt of \cite{Kaloper:2026llp} to reproduce the result of
\cite{Kaloper:2026auc} is inconsistent.

Below we present details of our analysis proving non-existence of the waves localized on the Chern--Simons interface.

\section{Maxwell Electrodynamics Interacting with Chern-Simons Interface}

\subsection{Problem statement. }

The author of \cite{Kaloper:2026auc} studied the system described by the sum of the standard Maxwell action and interface action given by the Chern--Simons 4--form multiplied by Heaviside step  function of $z=x^3$ coordinate,
\begin{equation}
S = -\frac{1}{4}\int d^4x F_{\mu\nu}F^{\mu\nu} - \frac{1}{8}\int d^4x \theta(x)\epsilon^{\mu\nu\lambda\rho}F_{\mu\nu}F_{\lambda\rho}\; ,
\label{eq:action}
\end{equation}
where $F_{\mu\nu}=\partial_\mu A_\nu - \partial_\nu A_\mu$,  $\theta(x) = \theta_- + \Delta\theta \Theta(z)$ with constant $\theta_-$ and $ \Delta\theta$, and $\Theta(z)$ is the Heaviside step function. Actually, the constant $\theta_-$ does not conribute to the equations of motion following from this action,
\begin{equation}\label{eq:eom}
\partial_\mu F^{\mu\lambda} + \frac{1}{2}\Delta\theta \delta(z)\epsilon^{\mu\lambda\sigma\rho}F_{\sigma\rho}\delta_\mu^3 = 0,
\end{equation}
so that we could set it to zero without loss of generality.

In three--vector notation $\vec{E} = (E_i)= (F^{0i})$, $B_i=\frac{1}{2}\epsilon_{ijk}F^{jk}$,   Eqs.~\eqref{eq:eom} and the Bianchi identities split into
\begin{eqnarray}
&& \nabla \cdot \vec{E} = -\Delta\theta \delta(z) B_z, \label{eq:gauss}\\
&& \nabla \times \vec{B} - \partial_t \vec{E} = \Delta\theta \delta(z) (\hat{z} \times \vec{E})\; , \label{eq:ampere}\\
&& \nabla \cdot \vec{B} = 0\; ,   \label{eq:divB=0}\\ && \nabla \times \vec{E} + \partial_t \vec{B} = 0\; .   \label{eq:dtB+=0}
\end{eqnarray}
Here and below  $\hat{x}$, $\hat{y}$, $\hat{z}$ are unity vectors in $x$, $y$ and $z$ directions.

As usually, and also following \cite{Kaloper:2026auc}, we will consider the equations below and above the interface plane (situated at $z=0$), which are homogeneous, and account for the effect of the sources in Eqs. \eqref{eq:gauss} and  \eqref{eq:ampere}  by imposing the matching conditions which are   \cite{Kaloper:2026auc}
\begin{align}
\vec{E}_\parallel(0^+) = \vec{E}_\parallel(0^-) &=: \vec{E}_\parallel (0), \qquad   \label{eq:bc3} \\  B_z(0^+) = B_z(0^-) &=: B_z(0),  \qquad  \label{eq:bc3+} \\
E_z(0^+) - E_z(0^-) &= -\Delta\theta B_z(0), \qquad \label{eq:bc1}\\
\vec{B}_\parallel(0^+) - \vec{B}_\parallel(0^-) &= \Delta\theta \, \vec{E}_\parallel(0)\; .  \qquad \label{eq:bc2}
\end{align}
Here
$\vec{E}_\parallel =E_x \hat{x} + E_y \hat{y}= E^L \vec{e}_R + E^R \vec{e}_L$ and $\vec{B}_\parallel =B_x \hat{x} + B_y \hat{y}= B^L \vec{e}_R + B^R \vec{e}_L$ are projections of electric and magnetic fields on the plane and we have introduced complex components of vectors, e.g. $E^L=\frac 1 {\sqrt{2}} (E_x + iE_y)=(E^R)^*$, and complex vectors
$\vec{e}_L=\frac 1 {\sqrt{2}}  (\hat{x}+i\hat{y})= (\vec{e}_R)^*$ obeying $\vec{e}_L\cdot  \vec{e}_R=1$. These latter are eigenvectors of the rotation operator $\hat{z}\times$, with eigenvalues $-i$ and $+i$, respectively.

Away from the wall ($z \neq 0$), the fields satisfy the standard vacuum wave equations,
\begin{align}\label{eq:wave}
& (\partial_t^2 - \vec{\partial}_{\vec{x}}^2) \vec{B}_{+} =0 \, , \quad   (\partial_t^2 - \vec{\partial}_{\vec{x}}^2) \vec{E}_{+} =0 \, , \quad z> 0\, , \quad \nonumber \\
& (\partial_t^2 - \vec{\partial}_{\vec{x}}^2) \vec{B}_{-} =0 \, , \quad   (\partial_t^2 - \vec{\partial}_{\vec{x}}^2) \vec{E}_{-} =0 \, , \quad z< 0\, , \quad
\end{align}
where  the sign subscripts, $+$ and $-$, corresponds to the branches of a solution defined for $z > 0$  and $z < 0$, respectively. Following \cite{Kaloper:2026auc}, let us consider the ansatz for electromagnetic waves parallel to the interface plane and localized on this plane, i.e. the waves which  decay exponentially with distance from the wall $z=0$,
\begin{eqnarray}
\vec{E}_\pm(x) = \vec{{\cal E}}_\pm\,  e^{\mp \kappa z}\, e^{i(\vec{k}_{\parallel}\cdot \vec{x}_{\parallel}-\omega t)}, \qquad \nonumber \\ \vec{B}_\pm(x) = \vec{{\cal B}}_\pm\, e^{\mp \kappa z} e^{i(\vec{k}_{\parallel}\cdot \vec{x}_{\parallel}-\omega t)}\; .  \qquad
\label{eq:ansatz}
\end{eqnarray}
The wave equations \eqref{eq:wave} impose the dispersion relation
\bea\label{k2=}
& \kappa^2 = \vec{k}_\parallel^2 - \omega^2\; .
\eea
\subsection{Non-existence of the waves localized on the Chern-Simons interface}
Further studying the  Maxwell equations \eqref{eq:gauss}--\eqref{eq:dtB+=0} with the above ansatz \eqref{eq:ansatz}, essentially as  in \cite{Kaloper:2026auc}, one arrives at the following relation between magnetic and electric amplitudes of the waves \be \label{vBpm=}\vec{{\cal B}}_\pm = \frac{1}{\omega}(\vec{k}_\parallel \mp i\kappa\hat{z}) \times \vec{{\cal E}}_\pm \; . \qquad \ee  The components of this equation which are  tangential to the interface are the same as  Eqs. (15) of \cite{Kaloper:2026auc}, namely
\bea\label{eq:cBLpm=}
{\cal B}^L_\pm =\frac 1 \omega \, \left[(\vec{k}_{\parallel} \times \vec{{\cal E}}_{\pm})^L\mp \kappa {\cal E}^L\right] \; , \qquad \\ \label{eq:cBRpm=}
{\cal B}^R_\pm =\frac 1 \omega \, \left[(\vec{k}_{\parallel} \times \vec{{\cal E}}_{\pm})^R\pm \kappa {\cal E}^R\right] \; , \qquad
\eea
where continuity of the component of electric field orthogonal to interface, Eq. \eqref{eq:bc3}, is taken into account, $ {\cal E}_+^{L,R}= {\cal E}^{L,R}_-=: {\cal E}^R$.

Up to this point we agree with \cite{Kaloper:2026auc}. However, the assertion above Eq. (16) of that  is incorrect. It states that {\it ''the terms $\propto$ $(\vec{k}_{\parallel}\times{\cal E})$ cancel in the difference of the magnetic fields across the interface''}, while this is not the case due to the discontinuity of the $z$--component of electric field on the interface, Eq. \eqref{eq:bc1}, which implies ${\cal E}_{z+} -{\cal E}_{z-}=-\Delta\theta\, {\cal B}_{z}$.
The difference of the magnetic fields across the interface calculated from Eqs. \eqref{eq:cBLpm=} and
\eqref{eq:cBRpm=} with discontinuity \eqref{eq:bc1} taking into account is given by
\bea\label{eq:cBL+--=}
& {\cal B}^L_+ - {\cal B}^L_-  =-  \frac {2\kappa} {\omega} \,  {\cal E}^L - \frac {i\Delta \theta} {\omega} \, k_L {\cal B}_z  \; , \qquad \nonumber \\ 
& {\cal B}^R_+ - {\cal B}^R_-  =\;\frac {2\kappa} {\omega} \,  {\cal E}^R + \frac {i\Delta \theta} {\omega} \, k_R {\cal B}_z  \; . \qquad \\ \nonumber
\eea
These differ from Eqs. (16) of \cite{Kaloper:2026auc} by the second terms $- \frac {i\Delta \theta} {\omega} \, k_R {\cal B}_z$ and $+ \frac {i\Delta \theta} {\omega} \, k_L {\cal B}_z$ which were missed there.

The study of the set of equations including the corrected versions of Eqs. (16) of \cite{Kaloper:2026auc}, given in our Eqs. \eqref{eq:cBL+--=}, shows that its consistency does not allow the wave solutions with $\kappa\not =0$.

Indeed, from the matching conditions \eqref{eq:bc2} we read that
\bea\label{cB+-cB-=DthcE}
& {\cal B}_+^{L, R} - {\cal B}_-^{L, R} = \Delta\theta\, {\cal E}^{L, R}\; .
\eea
Substituting these relations in  Eqs. \eqref{eq:cBL+--=}
one obtains  two equations for three variables ${\cal E}^{L,R}$ and ${\cal B}_z$,
\begin{eqnarray}
(2\kappa + \omega\Delta\theta) {\cal E}^L + i k_L \Delta\theta {\cal B}_z &= 0\, , \label{eq:sysL} \\
(2\kappa - \omega\Delta\theta) {\cal E}^R + i k_R \Delta\theta {\cal B}_z &= 0\, . \label{eq:sysR}
\end{eqnarray}
The third equation for these variables,
\begin{eqnarray}
- i k_R {\cal E}^L + i k_L {\cal E}^R  +\omega \,  {\cal B}_z &= 0\, , \label{eq:sysz}
\end{eqnarray}
is given by the $z$-component of \eqref{eq:cBLpm=}.

Thus we have a linear system of three equations for three variables  which has a nonvanishing solution if its determinant vanishes,
\begin{widetext}
\be
 \left| \begin{matrix}
2\kappa + \omega\Delta\theta & 0 & -i k_L \Delta\theta \\[0.1cm]
0 & 2\kappa - \omega\Delta\theta & -i k_R \Delta\theta \\[0.1cm]
-i k_R & + i k_L & +\omega
\end{matrix}\right|\, =  \omega (4\kappa^2 - \omega^2(\Delta \theta)^2+2k_Lk_R (\Delta \theta)^2)=0 \; .
\ee
Using the dispersion relation \eqref{k2=}, this equation reduces to
\bea
& \omega \kappa^2 (4+(\Delta \theta)^2)=0\;
\eea
\end{widetext}
which makes manifest that the nonvanishing solution requires either $\kappa=0$, so that the waves do  not decay exponentially with distance from the interface and hence are not localized on the latter, or $\omega=0$ so that the configuration is static.

The above arguments do not exclude the existence of static configurations localized on the interface.
However, the further study of the complete set of Maxwell equations \eqref{eq:gauss}--\eqref{eq:dtB+=0} with the ansatz \eqref{eq:ansatz} for $\omega=0$ shows that such static solutions exist only in the trivial case of $\kappa=0$.

\bigskip

\section{On arguments in \cite{Kaloper:2026llp} meant to support the conclusion of \cite{Kaloper:2026auc}. }

In Sec.~3 of  \cite{Kaloper:2026llp} the author provided an alternative argument to justify the existence of the plane wave solutions of \eqref{eq:eom} localized on the interface. It is based on a kind of analytic continuation whose consistency is doubtful.

In  Sec.~3$^{\cite{Kaloper:2026llp}}$ the author of {\cite{Kaloper:2026llp}} starts from writing Eq. \eqref{eq:eom} for the potential in the gauge  $\partial_\mu A^\mu=0=A^0$. With the ansatz describing the waves in the direction $z$ transverse to the interface and with amplitude dependent on $z$ coordinate only,
\be
A^\mu (t,z) =  e^{-i\omega t }\, (0,A^x(z), A^y(z), 0)\; ,
\ee
 Eq. \eqref{eq:eom} acquires the form
\be\label{eq:eomAxy}
(\partial_z^2+\omega^2) A^{x,y} = \pm  i \, \omega \Delta \theta \delta(z) A^{y,x}\; ,
\ee
or, in terms of the complex combinations \footnote{In this definition we use subscripts $_{[\pm]}$ instead of just  $_{\pm}$ in  {\cite{Kaloper:2026llp}}, to prevent confusion with $_{\pm}$ in the previous section and in \cite{Kaloper:2026auc} which mark the functions defining the solutions for $z>0$ and $z<0$, respectively. }
$A_{[\pm]}= \frac 1 {\sqrt{2}} \, (A^x\pm i A^y)$,
\be\label{ddzA+om2A=}
(\partial_z^2+\omega^2) A_{[\pm]}(z) =\pm   \omega \Delta \theta \delta(z) A_{[\pm]}(z)\; .
\ee
If the waves localized on the interface claimed in \cite{Kaloper:2026auc} existed, this would be reflected by
solutions of these equations which decay exponentially with distance $|z|$ from the interface.  This however cannot be obtained by the standard methods.

Indeed, as it was noted in \cite{Kaloper:2026auc}, Eq.
\eqref{ddzA+om2A=} is equivalent to the integral Lippmann--Schwinger equation \footnote{Let us recall that $\partial_z^2 |z-z^\prime| = 2\delta (z-z^\prime)$ so that $(\partial_z^2-\omega^2) e^{i\omega |z-z^\prime|}=2i\omega  \delta (z-z^\prime)$.}
\bea\label{eq:L-S}
 A_{[\pm]}(z) = {\cal A}_0 e^{i\omega z} - \frac i {2\omega}\, \int\limits_{-\infty}^{\infty}dz^\prime e^{i\omega |z-z^\prime|}V_{\pm}(z^\prime)  A_{[\pm]}(z^\prime)\; , \qquad \nonumber \\ V_{\pm}(z^\prime)=\pm  \Delta\theta\,  \omega \, \delta(z^\prime)\; . \qquad
\eea
The constant ${\cal A}_0$ is set to unity in  {\cite{Kaloper:2026llp}} (see Eq. (16)$^{\cite{Kaloper:2026llp}}$) and, for simplicity we will also set $ {\cal A}_0=1$ in the discussion below.

The straightforward calculation of the integral in the r.h.s. gives the term $\mp \frac i 2 \Delta \theta e^{i\omega |z|}  A_{[\pm]}(0)$ so that it is easy to find algebraically the expression for $ A_{[\pm]}(0)$ and then the solution of  \eqref{eq:L-S}. It is given by
\bea
 A_{[\pm]}(z) = e^{i\omega z} \mp  \frac i 2 \frac {\Delta\theta }{1\pm \frac i 2 \Delta \theta }e^{i\omega |z|}
  \nonumber \\ = \frac { e^{i\omega z} \mp \Theta(-z) \sin \omega z }{1\pm \frac i 2 \Delta \theta }\eea
and no traces of  waves localized on the interface is seen in this solution.

In desire to obtain the waves localized on the interface, as claimed \cite{Kaloper:2026auc}, in \cite{Kaloper:2026llp} the author  proposed the following procedure: to analytically continue $\omega^2$ in the l.h.s. of Eq. \eqref{ddzA+om2A=} to negative values, first denoting it by $\lambda$, while keeping
untouched  $\omega$ in the r.h.s. of \eqref{ddzA+om2A=}. This is tantamount to saying that Eq. \eqref{ddzA+om2A=} is replaced in \cite{Kaloper:2026llp} by
\be\label{ddzA+lA=}
(\partial_z^2+\lambda) A_{[\pm]}(z) =\pm   \omega \Delta \theta \delta(z) A_{[\pm]}(z)\; .
\ee
with setting then $\lambda=-\kappa^2$.

Clearly such equation has solutions exponentially decaying with distance from interface, $|z|$. However, the consistency of its derivation from the original Eq. \eqref{ddzA+om2A=} was looking doubtful.
Furthermore, our study in Sec. II actually provides a factual proof that it is inconsistent. Indeed,  this procedure, if accepted, would reproduce a particular class of the waves localized on the interface which do not exist as we have shown in Sec. II.

\section{Conclusion}

In this note we have shown that the wave  solutions localised on Chern-Simons interface, claimed to exist in \cite{Kaloper:2026auc}, actually do not exist. The reason for the erroneous conclusion in \cite{Kaloper:2026auc} has  been followed to erroneous  Eq. (16) of this paper, where some terms are missed.

The  waves parallel to interface do exist in the system introduced in \cite{Kaloper:2026auc}, but they are not decaying exponentially with the distance from the interface plane. So that they are de-localized, but, if
${\cal B}_z\not=0$, have a jump in their amplitudes when passing the interface.

\acknowledgments

The authors are
thankful to Dima Sorokin for reading this manuscript and for relevant suggestions, including to
write first a message to the author of \cite{Kaloper:2026auc} (no response, however, has been received).
The work of I.B. was supported in part by the Spanish  MCI, AEI and FEDER (ERDF EU) grant  PID2024-155685NB-C21``Gravity, Supergravity and Superstrings'' (GRASS) and by the Basque Government Grant ''Gravitation, Cosmology and Fundamental Physics'' IT1977-26.

\bigskip


\end{document}